\RequirePackage{fix-cm}
\documentclass[smallextended]{svjour3}       % onecolumn (second format)
\smartqed  % flush right qed marks, e.g. at end of proof
\usepackage{graphicx}
\usepackage{xcolor}
\usepackage{listings}
\usepackage{comment}
\usepackage{hyperref}
\begin{document}

\title{Viewing Computer Science through Citation Analysis}

\thanks{Research and development reported in this publication was partially supported by funds from the National Institute on Drug Abuse, National Institutes of Health, US Department of Health and Human Services, under Contract No HHSN271201800040C (N44DA-18-1216). TW is supported by the Grainger Foundation.}
\subtitle{Salton and Bergmark Redux}

\author{Sitaram Devarakonda  \and
	Dmitriy Korobskiy \and
        Tandy Warnow \and
        George Chacko }


\institute{Sitaram Devarakonda,  \at
              Netelabs, NET ESolutions Corporation, McLean, VA \\
              \email{sitaram@nete.com}           %  \\
           \and
              Dmitriy Korobskiy \at
              Netelabs, NET ESolutions Corporation, McLean, VA  \\
              \email{dk@nete.com}
           \and
           Tandy Warnow \at
              Dept of Computer Science, University of Illinois Urbana-Champaign, Champaign IL \\
              \email{warnow@illinois.edu}
           \and
           George Chacko \at
              Netelabs, NET ESolutions Corporation, McLean, VA \\
              \email{netelabs@nwete.com}
}

\date{Received: date / Accepted: date}

\maketitle

\begin{abstract}
Computer science has experienced dramatic growth and diversification over the last twenty years. Towards a current understanding of the structure of this discipline, we analyze 
a cohort of the computer science literature using the DBLP database. For insight on the features of this cohort and the relationship within its components, we constructed article 
level clusters based on either direct citations or co-citations, and reconciled them to major and minor subject categories in the All Science Journal Classification (ASJC). 
We described complementary insights from clustering by direct citation and co-citation, and both point to the increase in computer science publications and their scope.
Our analysis shows cross-category clusters, some that interact with external fields, such as the biological sciences, while others remain inward looking.

\keywords{Bibliometrics \and Clustering \and Research Evaluation \and Computer Science \and DBLP}
\subclass{01A85 \and 01A90} %\and more}

\end{abstract}

\section{Introduction}
\label{intro}

Computer science, and its applications, has experienced rapid growth and diversification over the last twenty years. As observed in a 2017 US National Academies Report, ``A wide range of jobs in virtually all sectors demand computing skills to an unprecedented extent. And every academic discipline finds itself incorporating computing into its research and educational mission"~\cite{nas_2017}. More recently, the collective influence of the Internet of Things (IoT), `big' data, accessible cloud computing, and advances in artificial intelligence have been postulated as a recent driver for growth and evolution~\cite{siebel2019_digital}. Given this powerful influence, some understanding of the present state and structure of computer science and its relationship to other fields could inform planning and policy making at multiple levels from national level funding all the way down to faculty hiring strategy.  

In historical precedent, Salton and Bergmark conducted a study  in 1979 of the computer science literature (419 computer articles published in 1974,  and 3,812 references cited in these articles)~\cite{salton_citation_1979}. Noting that that the scientific literature serves a rich source of information to study the structure and historical development of a field, these authors described the global structure of computer science as comprising three main areas: (i) theoretical foundations, such as theory of computation, (ii) hardware and computer systems, such as architecture,  and (iii) software, such as programming systems.  Related areas noted were (a) mathematics of computing, such as numerical analysis, (b) special software topics, such as operating systems, (c) data management and database systems, (d) methodologies valid for multiple applications, such as algebraic manipulation, (e) computer applications, such as computer graphics, and  (f) non-technical aspects, such as computer education. 

Looking beyond this historical triad of theoretical foundations, hardware and computer systems, and software, the Computing Classification System (CCS) published by the Association for Computing Machinery now consists of 13 top-level areas that reflect a more current view of the field~\cite{acm_ref}. This classification also addresses relationships with other fields under the category of Applied Computing. However, an easy way to map scientific publications to the CCS, especially interdisciplinary articles or those from proximal fields, does not seem to be available. 

Other classification systems are available, such as the All Science Journal Classification (ASJC) developed and maintained by Scopus, the Web of Science research and categorical classifications from Clarivate Analytics, and the National Science Foundation classification system~\cite{nsf_classification,scopus_ref,wos_ref}. Scopus and the Web of Science also include comprehensive bibliographies with citation links and proprietary unique identifiers for publications. The Scopus All Science Journal Classification (ASJC), which we use in this study, is organized  into 4 subject areas, 27 major subject areas, and 334 minor subject areas. All three systems rely on applying one or more journal-derived labels to articles.  
A logical prediction, given the diversity of articles within journals, is limited specificity at the article level. 
Shu and colleagues recently noted in a comparative study of the Chinese Science Citation Database (CSCD)  and the Web of Science that 46\% of articles did not belong to the discipline of the journal they were published in~\cite{shu_comparing_2019}. 
Others have also discussed and critiqued disciplinary assignments using journal-based classifications~\cite{wang_large-scale_2016,perianes-rodriguez_comparison_2017}. 
Article classification systems have been constructed that escape some of the criticisms of journal-based classification~\cite{traag_louvain_2019,boyack_classification_2014,waltman_new_2012}, but do not seem to presently enjoy widespread use. 
 
In this study, we extend the work of Salton and Bergmark through a combination of article and journal approaches to study trends in computer science, while also considering connections to other fields, especially biology since approximately 42\% of Scopus is classified under the top level subject areas of Life Sciences and Health Sciences. As a source of computer  literature, we used DBLP, a reference bibliography for computer science~\cite{dblp_ref}. The DBLP bibliography covers publications from computer science and includes publications from hybrid fields, where they are considered pertinent to computer science research. We describe below the high-level interactions of computer science within the Physical Sciences, and with the Social Sciences, Life Sciences, and Health Sciences. 

\section{Materials and Methods}
\label{sec:methods}

 \emph{Overview} For the purpose of this study, our working definition of the computer science literature was all publications in the DBLP bibliography that (i) had a digital object identifier (DOI), and (ii) could also be matched to article identifiers in the Scopus bibliography. The underlying assumption is that records in DBLP are greatly enriched for computer science even while recognizing that coverage may be incomplete and possibly biased. Cross-matching DBLP publications to records in the Scopus abstract and citation database of peer-reviewed literature enabled us to harvest the richer links in Scopus as well as extract links to publications from other disciplines. This cross-matching to Scopus also allowed the use of both journal- and article-based classifications when clustering documents. We used DBLP articles from journals and conference proceedings to construct article clusters by using either direct citations or co-citations as links. We reconciled these clusters with the All Science Journal Classification (ASJC) developed and maintained by Scopus through a combination of automated and manual procedures, producing a dataset of  2,685,356 publications.

\begin{figure}[ht]
  \includegraphics[scale=0.75]{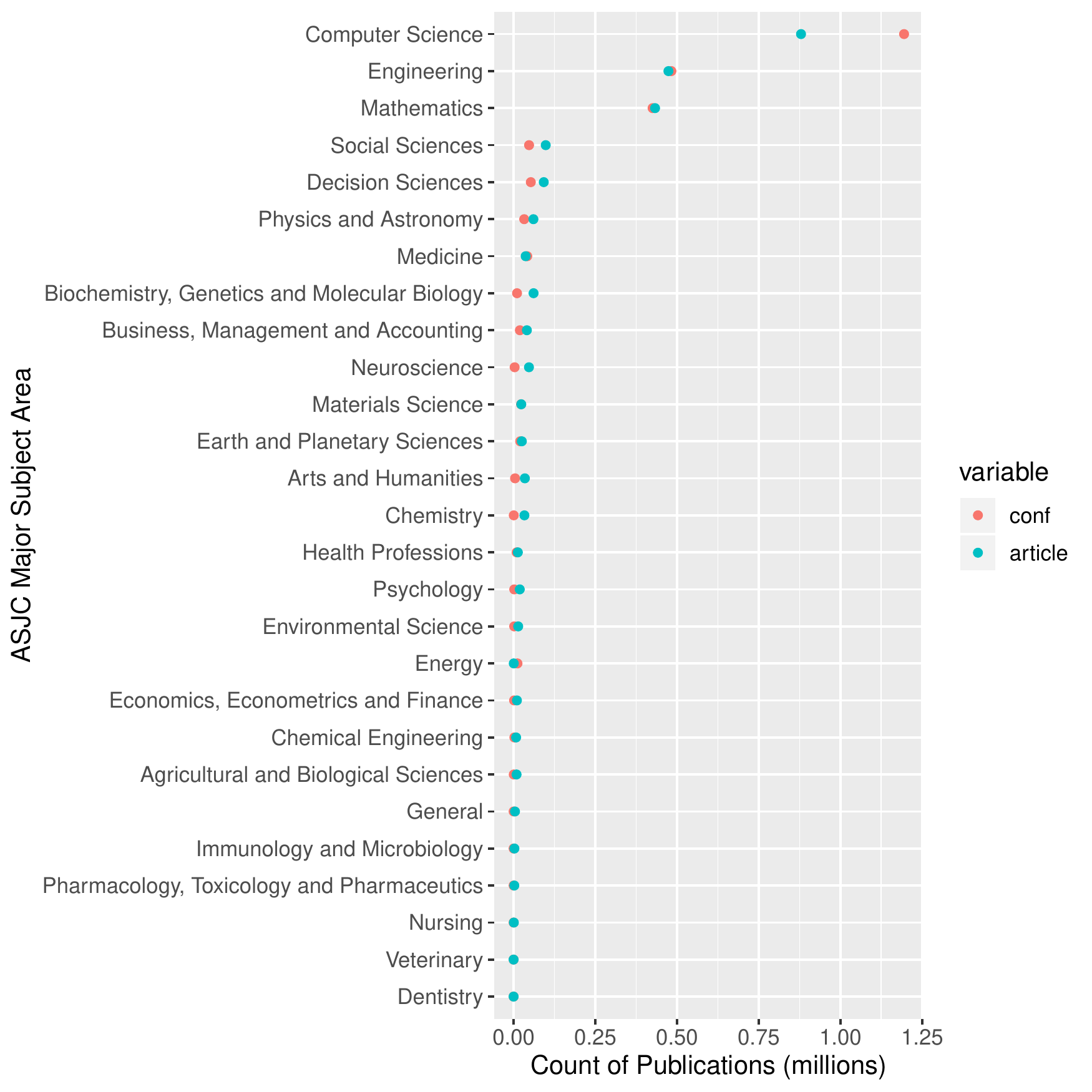}
\caption{Summary of DBLP data cross-matched with Scopus. 2,685,356 publications from DBLP were cross-matched with Scopus and then grouped by the 27 major subject areas in the ASJC (Scopus) classification. The largest number of publications are contributed by Computer Science; Engineering; Mathematics; and then by Social Sciences; Decision Sciences; Physics and Astronomy; Medicine; and Biochemistry, Genetics, and Molecular Biology. Publications were further annotated with respect to being either articles \emph{(ar)} or conference proceedings \emph{(cp)}. For this dataset, the major subject area of Computer Science with 1,194,623(cp) \& 879,396 contributed the most publications while Dentistry with 0 (cp) \& 1 (ar)) contributed the least.}
\label{fig:ar_cp_annotation}       % Give a unique label
\end{figure}

\emph{Data} A stable release of the DBLP computer science bibliography~\cite{dblp_ref} consisting of 7,079,994 records was downloaded as dblp-2018-08-01.xml.gz. 
Slightly over 95\% of the publications within were published after 1996. 
Publications were parsed from the XML source file and loaded into a PostgreSQL database. As part of implementing a larger data platform for research evaluation~\cite{GithubERNIE2019}, we have previously parsed the Scopus dataset, presently at over 88 million publications, into a custom schema in a PostgreSQL database. 
The total number of publications in Scopus labeled with major subject area Computer Science (in turn a subset of the Physical Sciences subject area) is 5,835,160. Records in the DBLP dataset were matched to Scopus identifiers using digital object identifiers (DOIs). This procedure resulted in a dataset of 2,685,356 DBLP publications with Scopus identifiers where 1,278,322 (47.6\%) were labeled as article and 1,407,034 (52.4\%) as conference proceedings.  References cited by these publications were then extracted from Scopus (7,129,006 records), resulting in a total of 8,000,411 
publications and references. 
We represented these data using a graph where the 8,000,411 nodes represent
publications and references  and the 44,296,381 undirected edges represent citations within the dataset. 
We refer to these data as the  \emph{comp} dataset (Fig.~\ref{fig:ar_cp_annotation}, Table~\ref{tab:comp}).  
 
\emph{Clustering} Clustering of publications is commonly accomplished through direct citation, bibliographic coupling, and co-citation, with direct citation being proposed as as the best approach to concentrate citation links~\cite{kessler_comparison_1965,klavans_which_2017}. Accordingly, we used direct citation links as the basis for cluster formation, and also co-citation to obtain an alternative view. In applying both clustering by direct citation and by co-citation, we attempted to consider, wherever possible, the criteria articulated by \v{S}ubelj, van Eck, and Waltman~\cite{subelj_clustering_2016} that (i) the largest cluster should be no more than 10 times the smallest one, (ii) small clusters should be eliminated, (iii) small changes and replicates should yield similar results (``stability"),  (iv) computing time should be minimized where possible, and (v) the clustering should seem reasonable on a qualitative level (``intuitive sensibility").

 \emph {Direct Citation} Graclus~\cite{graclus_2007} is a spectral graph clustering package that optimizes  various clustering criteria, including normalized cut, ratio cut, and ratio association, and that has previously been applied to citation data~\cite{subelj_clustering_2016}. 
We used v1.2 in our experiments. The \emph{comp} dataset was formatted as an undirected graph, stored in a file with a header line indicating the number of nodes and edges, and used as input to Graclus, which requires the number of clusters to be formed as an input parameter. In preliminary experiments, we varied the number of clusters to be formed between 10 and 50 clusters (data not shown).  At around 20 clusters, clusters size was relatively stable with the largest cluster containing roughly 10 times the number of nodes in the smallest one,
so that 20 clusters is a good choice with respect to the criteria specified in~\cite{subelj_clustering_2016}. Consequently, we used Graclus to generate 20 clusters, labeled 0-19 (Table~\ref{tab:graclus}), analogous to Level 1 of Waltman and van Eck's mapping of nearly 10 million publications but focused on the DBLP bibliography rather than a broader Web of Science sample~\cite{waltman_new_2012}. 

\begin{figure}[ht]
\centering
  \includegraphics[scale=0.75]{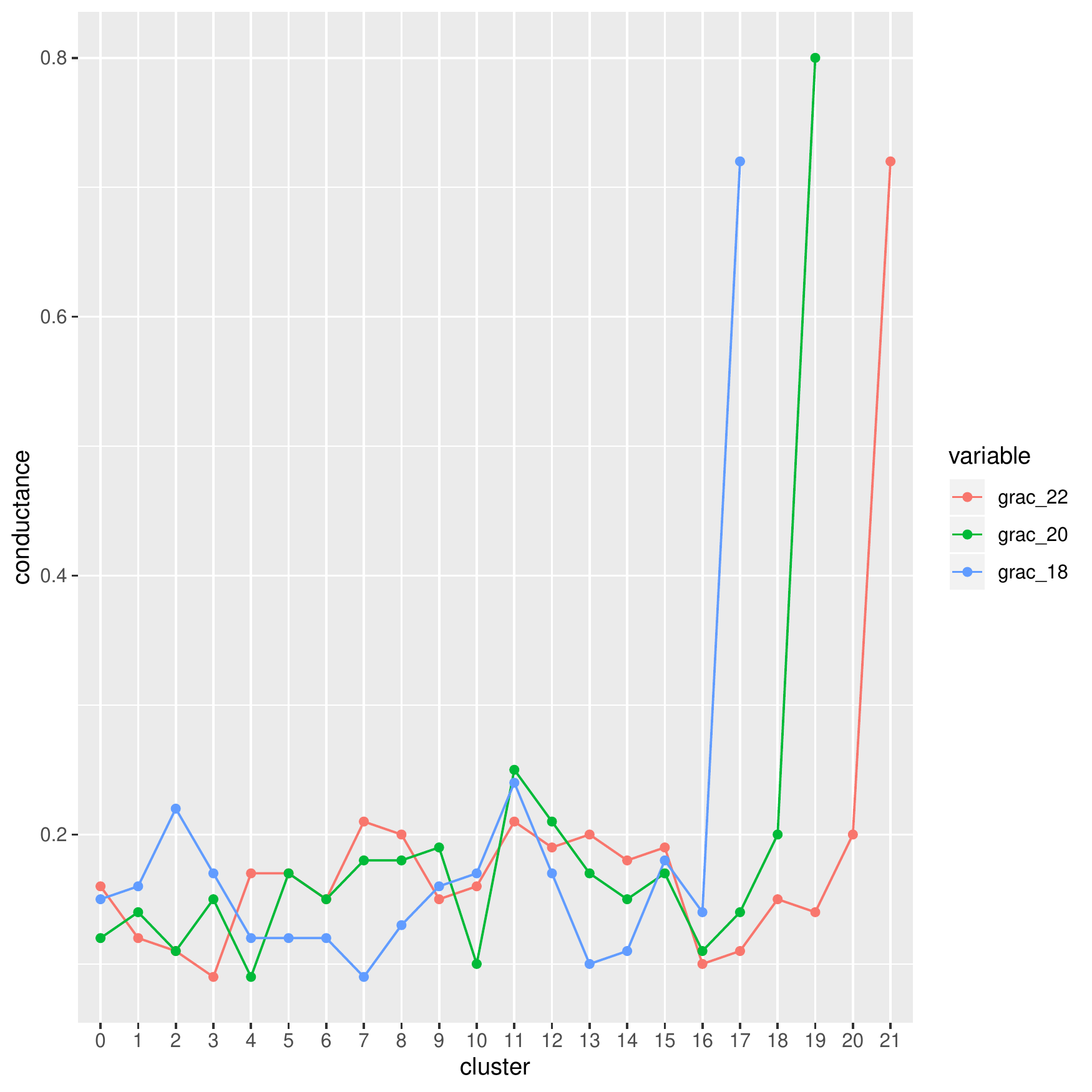}
\caption{Conductance Measurements of Clusters Generated by Graclus of the direct citation dataset.  2,685,356 DBLP publications, 7,129,006 cited references, and 44,296,381 citations were clustered using Graclus into 18 (grac\_18), 20 (grac\_20), or 22 (grac-22) clusters. Conductance, $\phi(S)$, was measured for these clusters considering only the edges between publications using the formula: $\phi(S)=|\partial(S)|/min(vol(S),2m-vol(S)$, where $\partial(S)$ is the boundary (number of edges leaving a set), vol(S) is volume of a set of vertices as the sum of the degrees of the vertices in a set, and $m$ is the number of undirected edges in a set~\cite{shun_parallel_2016}.}
\label{fig:graclus_comparison_conductance}       
\end{figure}

We also used conductance, as defined in Shun et al.~\cite{shun_parallel_2016}, to  evaluate clustering by direct citation (smaller is better), noting that conductance has been found to be a good metric for this purpose~\cite{emmons2016analysis,almeida_2012}. In our analysis, we saw that the last cluster (cluster 19) had a much larger conductance value than the other clusters and also had the smallest number of nodes. We then examined results obtained using Graclus with two other numbers of clusters (18 and 22), and in each case, the highest numbered cluster had the greatest conductance value and also the smallest number of nodes. These results suggest that Graclus produces a final cluster that effectively serves as a container for `left over publications' during the clustering procedure. Therefore, we limited our consideration of cluster 19 (the final of the twenty clusters)  when interpreting results.  The remaining 19 clusters had conductance values ranging from 0.09 to 0.25 with a median conductance of 0.15 (Fig.~\ref{fig:graclus_comparison_conductance}, Table~\ref{tab:comp}).

\emph{Co-Citation.} For an alternate view of these DBLP data, we constructed clusters using co-citation, the frequency with which a pair of articles is cited by other articles \cite{small_co-citation_1973,marshakova-shaikevich_co-citation_1973}. Co-citation, first described independently by Small and Marshakova in 1973 \cite{marshakova-shaikevich_co-citation_1973}, provides insight into the emergence of new ideas derived from the association of previously independent ones. Unlike clustering by direct citation, where every input publication is assigned to a cluster and every citation is weighted equally, the co-citation relationship between papers is weighted to represent the strength of the co-citation history. Because this produces a weighted graph, clustering methods that address weights are required. Clustering by co-citation also considers weak inter-cluster interactions that involve modifications to standard clustering approaches. \cite{boyack_cocitation_2010,boyack_improving_2013,small_structure_1974,small_clustering_1985}. 

We used a modification of variable level clustering combined with agglomerative clustering, an approach developed in 1985 by Small and Sweeney~\cite{small_clustering_1985} for co-citation analysis.
Variable level clustering involves applying a threshold (below which all edges in a graph are deleted) then iteratively selecting edges with the highest normalized co-citation value and extracting connected components from the graph as clusters for each edge in turn. Three parameters are needed: (i) a threshold or starting level based on a quantile of normalized co-citation frequency, (ii) a level increment, and (iii) a maximum cluster size. An issue is the generation of very large clusters by chaining via low edge weights.  Thus, at each iteration, any cluster exceeding the maximum cluster size  is returned to the process and a higher threshold is applied to break such clusters.

We first calculated the number of citations accumulated across Scopus for all 2,685,356 articles in \emph{comp}. After discarding those publications without any citations, we restricted further analysis to highly cited articles--those in the 90th percentile of citations or higher--resulting in a dataset of size 212,311. We then identified 4,318,305 publications in Scopus that cite these 212,311 articles from \emph{comp}. For each of the 4,318,305 citing publications in turn, all possible ${n \choose 2}$ reference pairs were generated from a publication's cited references, where $n$ is the number of references in a publication. The cited reference pairs this generated were then restricted to those in the set of 212,311 highly cited papers previously identified. A total of 46,463,117 unique co-cited pairs were thus obtained. 
The frequency of these co-cited pairs was then computed across the \emph{comp} dataset  and normalized using Salton's cosine formula~\cite{salton_citation_1979} to limit dominance by areas with high citation activity. These data were represented in a graph where each node was a publication and the weighted edge between the pair was the normalized co-citation frequency. \par 

\begin{figure}[ht]
\centering
\includegraphics[width=0.9\textwidth]{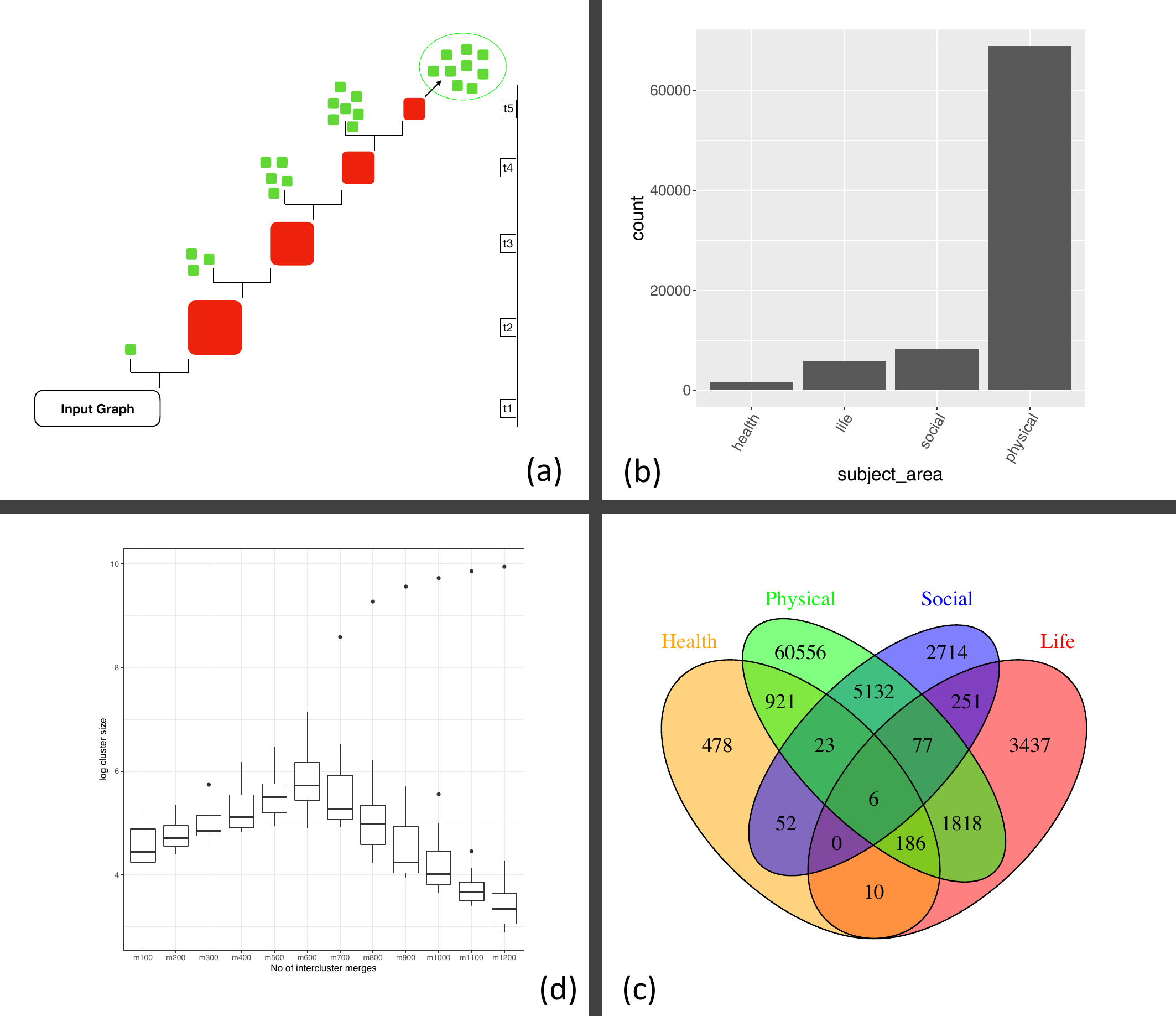}
\caption{Co-citation analysis.
(a) Schematic representation of variable clustering protocol modified from Small and Sweeney (1985) \cite{small_clustering_1985}. Three parameters are specified: (i) a threshold or starting level based on a quantile of normalized co-citation frequency, (ii) a level increment, and  (iii) a maximum cluster size. Input data is a set of co-cited publications with edge-weight defined by normalized co-citation frequencies. Green clusters are within the max cluster size. At the initial threshold, $t1$, a single cluster below the maximum cluster size, $mcs$ (green), along with one large cluster above it (red) are generated. As the threshold is incremented to $t2$, additional clusters of acceptable size is generated. The cascade continues to completion, which is defined by all clusters being of size less than or equal to the $mcs$. In this schematic, five rounds are adequate for the process to run to completion.
(b) The distribution of publications (using fractional counting) across four top-level ASJC subject areas after applying variable level clustering as in a)
(c) The Venn diagram of the fractional counting given in (b).
(d) The distribution of cluster sizes (logarithmic y-axis) as a function of the number of iterations of the agglomerative clustering technique; note that the largest cluster is extremely large when the number of iterations exceeds 600.
}
\label{fig:quad-chart}
\end{figure}

In our implementation of variable level clustering (Fig.~\ref{fig:quad-chart}(a)), 
we set initial parameter values as follows: the threshold  $t$  is initially set to  the median normalized co-citation frequency (quantile=0.5), increment $i = 0.1$, and maximum cluster size, $mcs=200$. Thus, at the start, all edges below the median normalized co-citation frequency were deleted. Clusters were formed by assembling connected components from each co-cited pair beginning with the heaviest edge weight. Clusters below size 100 were retained and any cluster larger than 200 nodes was carried over to the next round. The threshold, $t$, was then incremented by 0.1 and the process repeated while progressively incrementing $t$.  We used a bi-phasic approach where in which $t$ ranged from 0.5--0.9, after which $i$ was reduced to 0.01 for the range $0.9 \leq t \leq 0.99$. A final threshold of $t$=0.999 was applied to break the single remaining large cluster.  Using this approach, 22,232 clusters containing 84,591 nodes were generated, each containing less than 100 nodes. Clusters containing only two nodes were discarded, bringing the total number of clusters down to 10,298. The publications in these 10,298 clusters were overwhelmingly drawn from the Physical Sciences (one of the four top level categories in the Scopus ASJC classification), of which computer science is a sub-category (Fig.~\ref{fig:quad-chart}).

Agglomerative clustering was then performed on these 10,298 clusters to generate higher-order clusters. To focus on larger clusters, only those with at least 10 nodes were used as input. Briefly, each cluster was now treated as a node and the edge weight between two clusters was assigned to the maximum edge weight of all edges between the nodes in the two clusters. Edges were arranged in descending order. The first pair of clusters was merged and its edge weight with other interacting clusters was recalculated, again based on maximum edge weight. All edges were then re-ordered as before and the next pair of clusters was merged. The process was halted after 600 rounds to prevent large outlier clusters being generated (Fig.~\ref{fig:quad-chart}(d)).

\section*{Results}
\label{sec:results}

 In our high-level study of the structure of a twenty year cohort of the computer science literature, we chose to use both traditional journal-based and article-based approaches. Considering that traditional disciplines `may only partly reflect the actual organization of today's scientific research'~\cite{waltman_new_2012}, we constructed article clusters at high levels of aggregation using citations to examine the  computer science literature and also mapped these clusters to journal-based categories to take advantage of both article level and journal level approaches.

\emph{Data} The process of selection and matching resulted in a dataset of  publications (Materials and Methods). Of 4,291,130 DBLP publications with DOIs, only  2,685,356 had corresponding DOIs in Scopus. Of these roughly 2.68 million publications, approximately 2.07 million were assigned ASJC codes in Scopus corresponding to Computer Science (major subject area with 13 minor subject areas (Table~\ref{tab:comp}), with the balance of 610,000 publications non-exclusively shared between 26 different major subject areas ranging from Engineering (330,048) to Dentistry (Fig.~\ref{fig:ar_cp_annotation}). The set of 2.07 million publications classified under the major subject area Computer Science in Scopus spanned all 13 minor subject areas with Software at 30.3\%  being the largest component and Computer Science (miscellaneous) at 0.9\% the smallest. Publications in the \emph{comp} dataset labeled Theoretical Computer Science (409,082) are classified under the the major subject area of Mathematics rather than Computer Science (Table~\ref{tab:comp}). 

\iffalse
Overall, 38.86\% of the publications were assigned to one minor subject area, 78.1\% to two minor subject areas, 90.9\% to three minor subject areas, and 0.0003\% assigned to 8 different minor subject areas. 
\fi

\emph{Clustering by Direct Citation} We constructed article-level clusters of this computer science dataset at a sufficiently high level of aggregation to avoid cognitive challenge and cross-matched them to the Scopus ASJC classification. To focus on relatively high signal, we only considered Scopus ASJC minor subject area categories that accounted for at least 15\% of the publications in each cluster. 

\begin{figure}[ht]
\centering
  \includegraphics[scale=0.6]{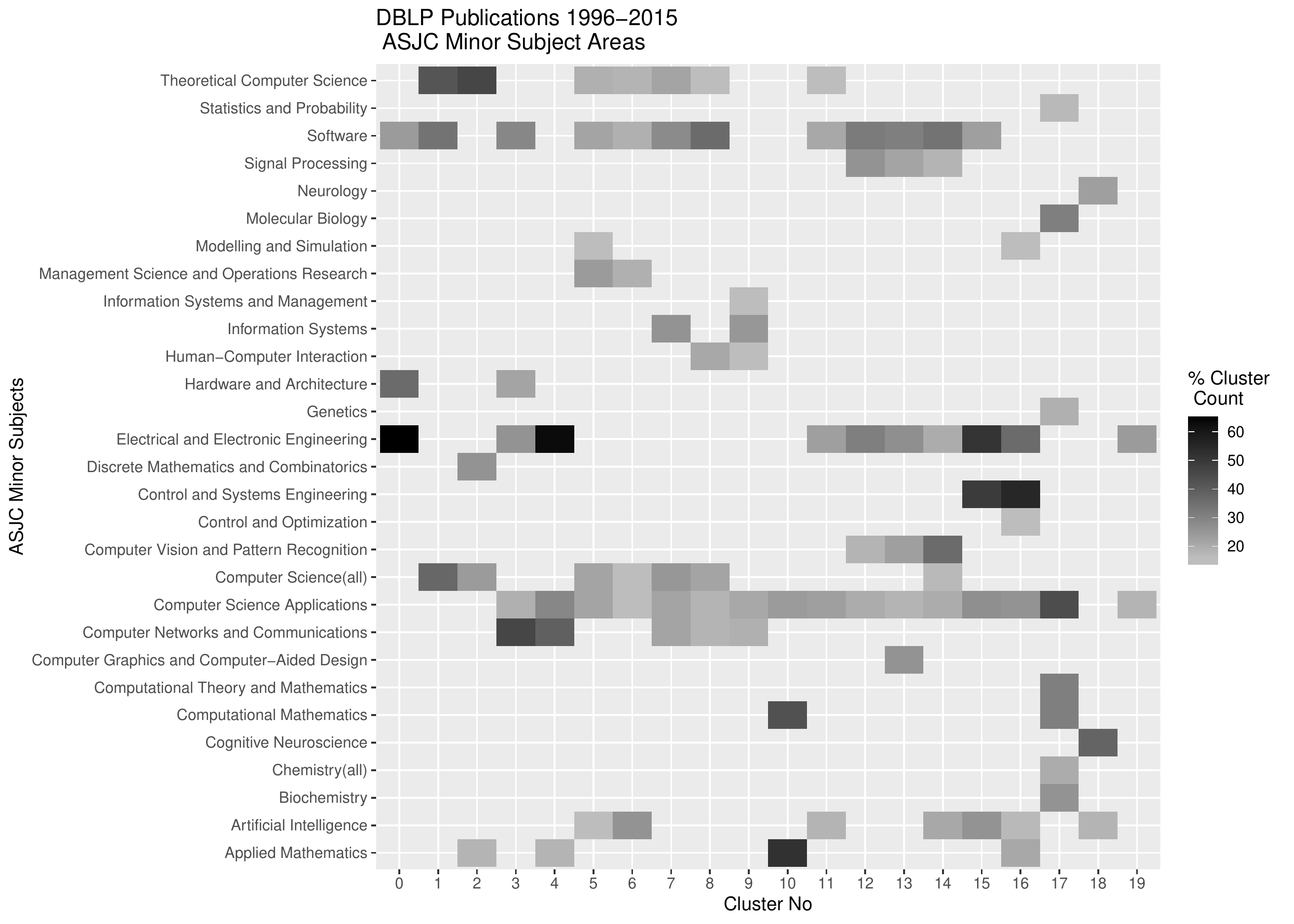}
\caption{Heat map for the clustering obtained by direct citation. The  y-axis (rows) correspond to topics, defined by Scopus characterizations, and the
x-axis (columns) represent the 20 different clusters.  
Each cluster is characterized by topics that label at least 15\% of the publications in the cluster.
 }
\label{fig:heatmap}       % Give a unique label
\end{figure}

Figure~\ref{fig:heatmap} permits examination of these data from two perspectives: (i) rows: the clusters that map to a given ASJC minor subject area and (ii) columns: ASJC minor subject areas that comprised at least 15\% of the publication in a cluster. Under these conditions of clustering and this threshold of 15\%, 31 of the 334 ASJC minor subject areas are represented. Unsurprisingly, the broad categories Computer Science Applications, Software, and Electrical and Electronic Engineering register in 16, 12, and 10 clusters respectively, while Artificial Intelligence mapped to 7 different clusters. At the other end of the range, 12 of the 31 ASJC minor subject areas were each detected only in a single cluster. 

From the alternate perspective (columns), Cluster 17 was the most diverse and contained publications annotated with 8 minor subject area labels: Biochemistry, Chemistry(all), Genetics, Molecular Biology,  Statistics \& Probability, Computational Theory \& Mathematics, and Computational Mathematics. Cluster 19 mapped to two areas but was excluded from qualitative analysis because of its high conductance value. 
Of the remaining clusters, Cluster 2 represents interactions between the four minor subject areas Theoretical Computer Science, Discrete Mathematics and Combinatorics, Applied Mathematics,  and the more generic Computer Science (all). Clusters 3 and 4 include Computer Networks and Communications, and Cluster 18 (Artificial Intelligence, Cognitive Neuroscience, and Neurology) and clusters 5--9 include Management Science, Operations Research, Information Systems, Modeling and Simulation, and Human-Computer Interaction.

These data suggest that fields central to computer science in 2019 (Salton and Bergmark's historical triad of hardware, software, and theory) are more likely to be found in multiple clusters than peripheral fields. A second inference is that, in some cases, journal-based classification and our article clusters align fairly well (Hardware and Architecture). 
A third inference is that the ASJC minor subject area ``Computer Applications" is relatively broad, and publications thus labeled are present at the $>=$ 15\% level in 16 out of 20 clusters.  Finally,   for this DBLP dataset, as we clustered it, interactions with fields outside computer science such as Biology  (i.e., Biochemistry, Neurology)  are  detected in two separate clusters. The first appears to be the interaction of Biochemistry, Molecular Biology, and Genetics with Statistics, Mathematics, and Computer Science, and the second is the interaction of Neurology, Cognitive Neuroscience, and Artificial Intelligence.

These clusters cannot be easily characterized by using the CCS classification. For example, we manually matched the top 25--50 most heavily cited publications in each cluster to corresponding categories in the CCS. This was feasible with clusters 0 and 1 mapped reasonably well to the top level categories Hardware and Computer Systems Organization, but in other cases, the top cited papers often derived from biology, yet biology was clearly not  representative of the majority of the nodes in these clusters. 

\begin{figure}[ht]
\centering
  \includegraphics[scale=0.6]{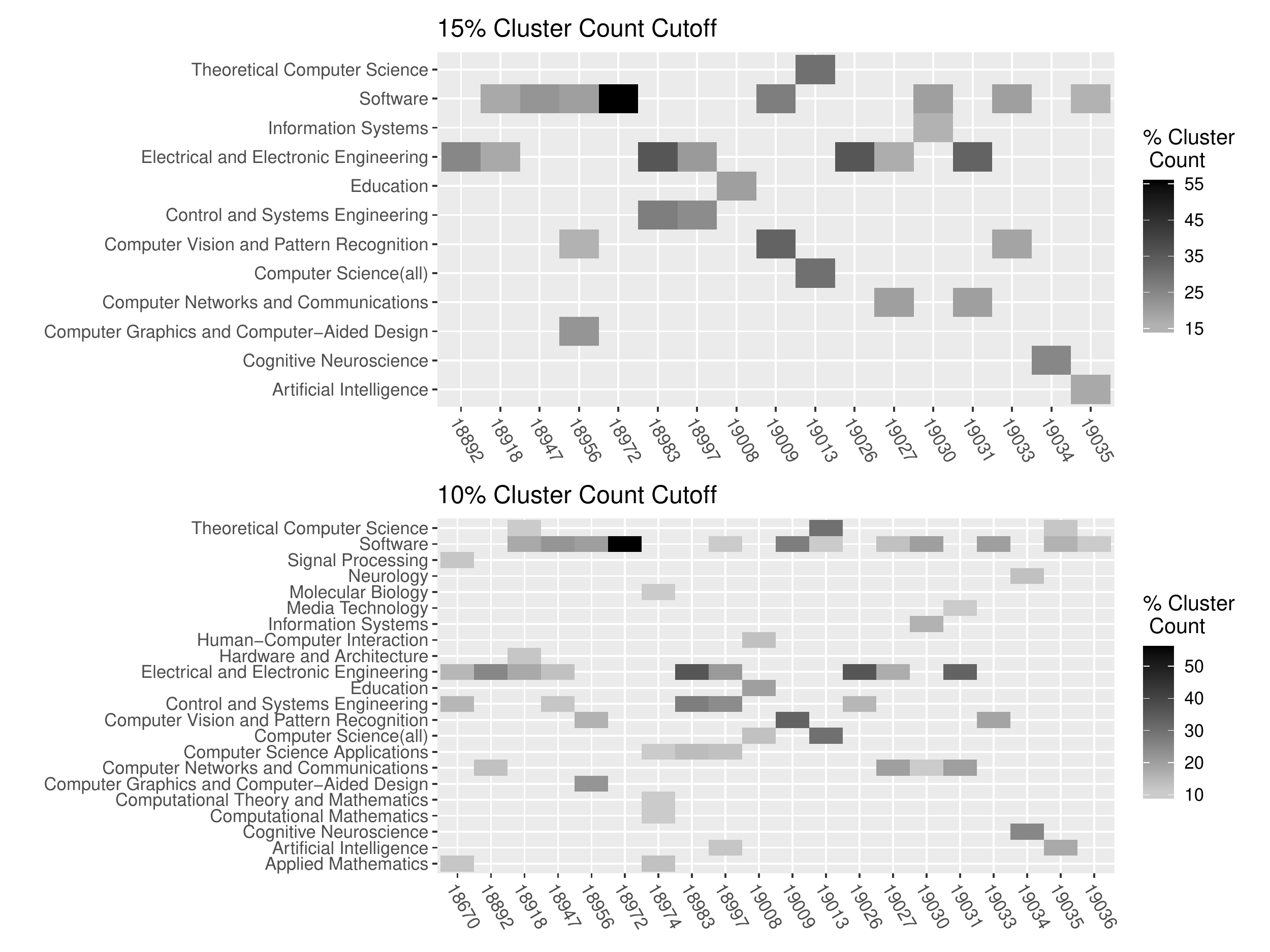}
\caption{Heat map for the clustering obtained by co-citation, using two thresholds for inclusion (top: 15\%, bottom: 10\%). The  y-axis (rows) correspond to topics, defined by Scopus characterizations, and the
x-axis (columns) represent the 20 different clusters.  
Each cluster is characterized by topics that label at least the required minimum percentage of publications in the cluster (top: 15\%, bottom: 10\%).   }
\label{fig:heatmap_cocit}       % Give a unique label
\end{figure}

\emph{Clustering by Co-citation.} For an alternate examination of the data, we used co-citation frequencies to cluster the DBLP dataset as described above (Materials and Methods). Figure \ref{fig:heatmap_cocit} shows a heatmap in which clusters constructed by co-citation are mapped to Scopus ASJC minor subject area labels, with the top subfigure showing results for those labels that account for at least 15\% of the publications in a cluster, and the bottom subfigure showing results for those labels that account for at least 10\% of the publications. 

At the threshold of 15\%, only 17 of 20 co-citation clusters mapped to at least one minor subject areas. At the 15\% threshold, no co-citation cluster maps to more than three minor subject areas, in contrast to a maximum of eight minor subject areas for clustering by direct citation (Fig. 3, Cluster 17), . The threshold had to be reduced to 10\% for all 20 co-citation clusters to map to at least one minor subject area. We interpret these data as indicative of broader clusters created by weaker linkages that accumulate during the agglomerative clustering phase~\cite{small_clustering_1985}. 

Clustering by co-citation begins, for each cluster, with a pair of nodes that nucleates its subsequent formation. Thus, we also designated the pair of nodes in a cluster with the strongest edge as its nucleating pair and labeled the cluster by manually labeling the nucleating pair of documents.  
These labels show a high degree of correspondence  to the ASJC minor subject area that accounts for the largest fraction of the nodes in a co-citation cluster (Table~\ref{tab:centroid_reconcile}). 
%The extent of this correspondence %between the manually applied label and the ASJC minor subject area 
Thus, using the nucleating pair as the basis for labeling co-citation clusters (as we generated them) may be valid, although its usefulness is likely to vary according to the data being examined; we also note that manual annotation (although beneficial) is not scalable. We provide the DOIs of these nucleating pairs matched to manually assigned labels for independent review (Table~\ref{tab:centroid}).

\begin{figure}[ht]
\centering
  \includegraphics[scale=0.6]{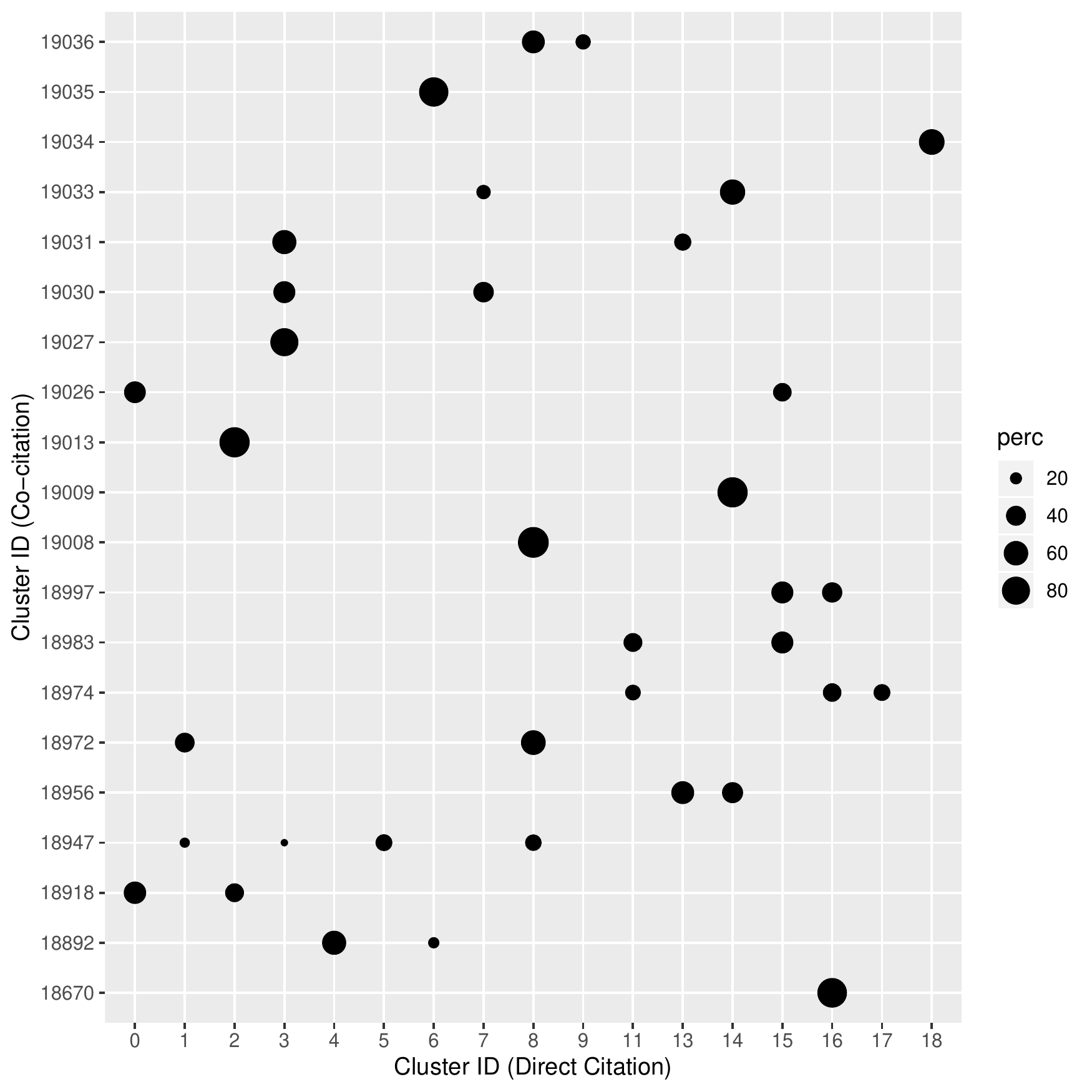}
\caption{Intersections between clusters generated using direct citation and co-citation features. x-axis: Clusters generated by Graclus number 0-19. y-axis: Cluster generated by variable level clustering and agglomerative clustering using a modification of Small and Sweeney (1985)~\cite{small_clustering_1985}. Point size (perc) is the percentage of a co-cited cluster that maps to a corresponding Graclus cluster. A minimum threshold of 15\% was set. Graclus cluster 19, the 20th cluster, did not map to any cluster on the y:axis.}
\label{fig:graclus_cocit_fig}       % Give a unique label
\end{figure}

To examine the correspondence between clusters generated by direct citation vs co-citation, we mapped the contents of these clusters to each other (Fig.~\ref{fig:graclus_cocit_fig}).
 At a threshold of 15\%, similar to other cross-matching, 90\% (18/20) of the co-cited clusters mapped to 1 or 2 direct citation clusters. This suggests that the majority of co-cited pairs tend to lie within the same cluster of publications linked by citations and,  by extension, tend towards disciplinarity rather than interdisciplinarity.

\section{Discussion} 
Considering expansion and diversification of the field of computer science, we revisited its characterization by Salton and Bergmark in 1979~\cite{salton_citation_1979}. In comparison to \cite{salton_citation_1979}, we analyzed considerably more data, 2.68 million publications versus 391 by using two bibliographic databases, DBLP and Scopus, consisting of  approximately 7 million and 88 million publications respectively. By linking the two datasets to harvest citation and metadata, we were able to construct article level clusters in two different ways and reconcile these clusters with the Scopus ASJC classification, which provides a journal level perspective.

Reconciling direct citation clusters to the Scopus ASJC classification yielded partially overlapping results that are consistent with the observation of Waltman and van Eck (2012)~\cite{waltman_new_2012} `that traditional disciplines such as those just mentioned only partly reflect the actual organization of today’s scientific research'. Of interest to us was the single obviously multidisciplinary cluster in which at least 15\% of its component publications were labeled with the ASJC minor subject areas Biochemistry, Chemistry, Computational Mathematics, Computational Theory and Mathematics, Computer Science Applications, Genetics, Molecular Biology, Statistics \& Probability. A second cluster mapped to Artificial Intelligence, Cognitive Neuroscience, and Neurology.  Both suggest collaboration between computer science and biology. At the opposite end of this spectrum is the cluster that maps to Hardware and Architecture, Electrical and Electronic Engineering, and Software. These data suggest that, at least from the perspective of a high level of aggregation, some subfields within computer science may be primarily inward looking (remain concerned with fundamental questions in computer science and electrical engineering), while others are more actively engaged with fields external to computer science. 

A comparison of the clusters generated using direct citation and co-citation shows interesting contrasts. The nodes in a co-citation cluster often map largely to one or two direct citation clusters: for example, 98\% of the nodes in co-citation cluster 19008 map to direct citation cluster 8, which in turn aligns with theory, software, applications, and networks. Conversely, co-citation cluster 18947 nucleated by a pair of articles in the Journal of Applied Mathematics and Computation is distributed between direct citation clusters 11, 16, and 17, effectively spanning Artificial Intelligence, Applied Mathematics, Biochemistry, Chemistry, Genetics, Theory, Software, Applications, and Statistics. 

The focus of this article was on high-level features, and a preference for simplicity and intuitiveness in the choice of methods. This preference may address questions  regarding generalizability and the specific choices we made, such as the (i) use of the DBLP dataset matched to Scopus, which may not capture all aspects of computer science and its interactions with other fields, (ii) the basis for clustering, and (iii) mapping the results of this clustering against a classification designed around journals rather than individual articles. We believe that an approach that combines journal-level  with article-level analyses is useful for studies of this kind.  

We speculate that biology, often dominant in bibliometric studies, is restricted to two clusters on account of (i) the focus of DBLP,  (ii) our use of normalized co-citations, and (ii) the threshold set for detection. Future research should, of course, include complementary investigations at finer levels of granularity and sensitivity using article-level and topic approaches that others have developed~\cite{glanzel_using_2017,boyack_clustering_2011,boyack_investigating_2017,boyack_classification_2014,sjogarde_granularity_2019,traag_louvain_2019}. We also refer readers to a related study of the computer science literature~\cite{chakraborty_role_2018} that is focused on evolving interdisciplinarity in computer science using data from Microsoft Academic Research and a 24 category classification of Computer Science.

In the 40 years since Salton and Bergmark's landmark paper, the field of computer science has not only expanded in volume, it has expanded in its interactions with other fields, and has also resulted in new disciplinary and interdisciplinary subfields. Furthermore, machine learning and data science, which build off computer science and statistics, are emerging as major fields that are driving innovation in industry and science, and research in these areas is increasingly being performed in fields external to computer science.While DBLP provides an insight into what is commonly accepted as computer science, additional evaluation of the broader literature that uses and develops computer science is needed to better assess the impact of computer science.

\clearpage

%%%%%%%%%
\section*{Figures}
%%%%%%%%%

\newpage

\newpage

\newpage

\newpage

\newpage

\newpage
\clearpage
\section*{Tables}

\begin{table}[ht]
\caption{Characterization of the \emph{comp} dataset relative to Scopus ASJC minor subject areas.}
\label{tab:comp}       
\begin{tabular}{lrccc}
\hline\noalign{\smallskip}
minor\_subject\_area & Percent of Publications \\
\noalign{\smallskip}\hline\noalign{\smallskip}
Software & 30.1 \\
Computer Science Applications & 25.0 \\ 
Computer Networks and Communications & 21.5 \\
Theoretical Computer Science & 19.6 \\ 
Computer Science(all) & 19.4 \\ 
Artificial Intelligence & 14.2 \\ 
Information Systems & 11.8 \\ 
Hardware and Architecture & 11.8 \\ 
Signal Processing & 9.8 \\ 
Computer Vision and Pattern Recognition & 9.4 \\ 
Computational Theory and Mathematics & 7.9 \\ 
Human-Computer Interaction & 7.4 \\ 
Computer Graphics and Computer-Aided Design & 5.2 \\ 
Computer Science (miscellaneous) & 0.9 \\ 
\noalign{\smallskip}\hline
\end{tabular}
\end{table}
\newpage

\begin{table}[ht]
\caption{Descriptive statistics of the 20 clusters produced from the \emph{dataset} using Graclus (direct citation).  The number of publications in each cluster, the conductance~\cite{shun_parallel_2016} of each cluster, the total number ASJC minor subject area labels assigned to publications in each cluster and the number of 
unique labels in each cluster are shown.}
\label{tab:graclus}       
\begin{tabular}{lrccc}
\hline\noalign{\smallskip}
Cluster & Publications & Conductance & Total ASJC Labels & Unique ASJC Labels\\
\noalign{\smallskip}\hline\noalign{\smallskip}
0 & 111,294 & 0.12 & 265,664 & 142 \\ 
1 & 117,057 & 0.14 & 246,960 & 166 \\ 
2 & 116,251 & 0.11 & 280,602 & 165 \\ 
3 & 353,366 & 0.15 & 881,693 & 200 \\ 
4 & 145,081 & 0.09 & 349,020 & 154 \\ 
5 & 92,097 & 0.17 & 248,168 & 186 \\ 
6 & 71,865 & 0.15 & 199,681 & 163 \\ 
7 & 179,927 & 0.18 & 465,181 & 202 \\ 
8 & 302,656 & 0.18 & 760,117 & 214 \\ 
9 & 69,520 & 0.19 & 197,031 & 174 \\ 
10 & 42,462 & 0.10 & 102,838 & 141 \\ 
11 & 448,030 & 0.25 & 1,229,061 & 224 \\ 
12 & 70,738 & 0.21 & 216,546 & 179 \\ 
13 & 105,232 & 0.17 & 289,318 & 187 \\ 
14 & 199,176 & 0.15 & 551,657 & 208 \\ 
15 & 64,384 & 0.17 & 195,679 & 167 \\ 
16 & 89,340 & 0.11 & 240,157 & 158 \\ 
17 & 50,817 & 0.14 & 181,531 & 179 \\ 
18 & 43,113 & 0.20 & 108,518 & 177 \\ 
19 & 12,615 & 0.80 & 36,583 & 229 \\ 
\noalign{\smallskip}\hline
\end{tabular}
\end{table}
\newpage

\begin{table}[ht]
\caption{\emph{Nucleating Co-citations.} For each of the 20 co-citation clusters, the pair with the strongest edge (greatest normalized co-citation frequency) is shown below along with a manually assigned label.}
\label{tab:centroid}       
\begin{tabular}{p{8 mm}p{42 mm}ll}
\hline\noalign{\smallskip}
Cluster & Nucleating Pair & NCF & Manual Label \\
\noalign{\smallskip}\hline
18670 & 10.1016/j.cam.2015.03.057 10.1016/j.sigpro.2015.10.009 & 0.92 & Dynamical Systems \\
18892 & 10.1287/msom.1080.0228 10.1287/msom.1060.0190 & 0.72 & Operations Research\\ 
18918 & 10.1007/s11128-010-0177-y 10.1007/s11128-013-0567-z & 0.83 & Image Processing\\ 
18947 & 10.1109/TASE.2011.2160452 10.1109/TASE.2011.2178023 & 0.77 & Robotics\\ 
18956 & 10.1109/ICCV.2017.32 10.1109/ICCV.2017.31 & 0.84 & Computer Vision\\ 
18972 & 10.1109/ASE.2013.6693094 10.1145/2568225.2568254 & 0.69 & Software \\ 
18974 & 10.1016/j.amc.2009.03.023 10.1016/j.amc.2010.07.064 & 0.54 & Applied Mathematics\\ 
18983 & 10.1137/110848864 10.1137/110848876 & 0.66 & Optimization\\ 
18997 & 10.1504/IJMIC.2014.065339 10.1504/IJMIC.2015.068871 & 0.91 & Chaotic Systems \\ 
19008 & 10.1016/j.chb.2008.12.013 10.1016/j.chb.2008.12.012 & 0.63 & Hum Comp Interaction\\ 
19009 & 10.1109/AVSS.2017.8078491 10.1109/ICCV.2017.206 & 0.78 & Artificial Intelligence\\ 
19013 & 10.1145/2508859.2516668 10.1007/978-3-642-42045-0\_15 & 0.88 & Security\\ 
19026 & 10.1145/2541940.2541942 10.1109/HPCA.2014.6835965 & 0.81 & Architecture\\ 
19027 & 10.1016/S0305-0548(03)00250-8 10.1016/j.ejor.2006.03.013 & 0.82 & Graph algorithms\\ 
19030 & 10.1109/ICDE.2008.4497474 10.14778/1687627.1687666 & 0.69 & Databases \\ 
19031 & 10.1109/TMM.2005.843347 10.1109/TCE.2005.1405724 & 0.55 & Networks\\ 
19033 & 10.1109/TIFS.2014.2327757 10.1109/TCYB.2014.2376934 & 0.78 & Facial Recognition\\ 
19034 & 10.1016/j.neuroimage.2013.05.018 10.1016/j.neuroimage.2014.06.016 & 0.63 & Neurology\\ 
19035 & 10.1002/int.21933 10.1002/int.21927 & 1.09 & Intelligent Systems\\ 
19036 & 10.1006/jsco.2000.0402 10.1006/jsco.2000.0403 & 0.73 & Comp Geometry\\ 
\noalign{\smallskip}\hline
\end{tabular}
\end{table}
\newpage

\begin{table}[ht]
\caption{\emph{Nucleating Co-citations} Manually assigned labels for nucleating co-cited pairs (Table~\ref{tab:centroid}) are matched to the ASJC minor subject area that constitutes the largest fraction (shown as percentage in parentheses) of all nodes in the cluster. In cases of ties, both minor subject areas are shown. Abbreviations: Electrical and Electronic Engineering (EEE); Control and Systems Engineering (CSE).}
\label{tab:centroid_reconcile}       
\begin{tabular}{p{8 mm}p{42 mm}l}
\hline\noalign{\smallskip}
Cluster & Label & Minor Subject Area \\
\noalign{\smallskip}\hline
18670 & Dynamical Systems & (i) EEE (14) (ii) CSE (14) \\
18892 & Operations Research & EEE (24) \\ 
18918 & Image Processing & (i) EEE (17) (ii) Software (17) \\ 
18947 & Robotics & Software (12) \\ 
18956 & Computer Vision & Computer Graphics and Computer-Aided Design (21)\\ 
18972 & Software & Software (55) \\ 
18974 & Applied Mathematics & Applied Mathematics (12)\\ 
18983 & Optimization & EEE (35) \\ 
18997 & Chaotic Systems & CSE (23)\\ 
19008 & Hum Comp Interaction & Education (19)\\ 
19009 & Artificial Intelligence & Computer Vision and Pattern Recognition (32)\\ 
19013 & Security & Computer Science(all) (29)\\ 
19026 & Architecture & EEE (35) \\ 
19027 & Graph algorithms & Computer Networks and Communications (19)\\ 
19030 & Databases & Software (19)\\ 
19031 & Networks & EEE (32)\\ 
19033 & Facial Recognition & Computer Vision and Pattern Recognition (18)\\ 
19034 & Neurology & Neurology (12)\\ 
19035 & Intelligent Systems & Artificial Intelligence (17)\\ 
19036 & Comp Geometry & Software (10)\\ 
\noalign{\smallskip}\hline
\end{tabular}
\end{table}
\newpage

\begin{acknowledgements} The authors thank Henry Small for very helpful discussions. Research and development reported in this publication was partially supported by funds from the National Institute on Drug Abuse, National Institutes of Health, US Department of Health and Human Services, under Contract No HHSN271201800040C (N44DA-18-1216). TW is supported by the Grainger Foundation. Citation data used in this paper relied on Scopus (Elsevier Inc.) as implemented in the ERNIE project (Korobskiy et al., 2019), which is collaborative between NET ESolutions Corporation and Elsevier Inc.
\end{acknowledgements}

 \section*{Conflict of interest}
The authors declare that they have no conflicts of interest. The content of this publication is solely the responsibility of the authors and does not necessarily represent the official views of the National Institutes of Health, NET ESolutions Corporation, or Elsevier Inc.

\bibliographystyle{spmpsci}      % mathematics and physical sciences
\bibliography{comp}   % name your BibTeX data base

\begin{thebibliography}{10}
\providecommand{\url}[1]{{#1}}
\providecommand{\urlprefix}{URL }
\expandafter\ifx\csname urlstyle\endcsname\relax
  \providecommand{\doi}[1]{DOI~\discretionary{}{}{}#1}\else
  \providecommand{\doi}{DOI~\discretionary{}{}{}\begingroup
  \urlstyle{rm}\Url}\fi

\bibitem{almeida_2012}
Almeida, H., Guedes, D., Meira~Jr, W., Zaki, M.: Towards a better quality
  metric for graph cluster evaluation.
\newblock Journal of Information and Data Management {(JIDM)} \textbf{3},
  378--393 (2012)

\bibitem{acm_ref}
{Association for Computing Machinery}: {Computing Classification System}
  (2012).
\newblock \urlprefix\url{https://dl.acm.org/ccs/ccs.cfm}.
\newblock Accessed June 2019

\bibitem{boyack_cocitation_2010}
Boyack, K., Klavans, R.: Co-citation analysis, bibliographic coupling, and
  direct citation: {Which} citation approach represents the research front most
  accurately?
\newblock Journal of the American Society for Information Science and
  Technology \textbf{61}(12), 2389--2404 (2010).
\newblock \doi{10.1002/asi.21419}

\bibitem{boyack_investigating_2017}
Boyack, K.W.: Investigating the effect of global data on topic detection.
\newblock Scientometrics \textbf{111}(2), 999--1015 (2017).
\newblock \doi{10.1007/s11192-017-2297-y}.
\newblock \urlprefix\url{https://doi.org/10.1007/s11192-017-2297-y}

\bibitem{boyack_clustering_2011}
Boyack, K.W., Newman, D., Duhon, R.J., Klavans, R., Patek, M., Biberstine,
  J.R., Schijvenaars, B., Skupin, A., Ma, N., Börner, K.: Clustering {More}
  than {Two} {Million} {Biomedical} {Publications}: {Comparing} the
  {Accuracies} of {Nine} {Text}-{Based} {Similarity} {Approaches}.
\newblock PLOS ONE \textbf{6}(3), e18029 (2011).
\newblock \doi{10.1371/journal.pone.0018029}.
\newblock
  \urlprefix\url{https://journals.plos.org/plosone/article?id=10.1371/journal.pone.0018029}

\bibitem{boyack_classification_2014}
Boyack, K.W., Patek, M., Ungar, L.H., Yoon, P., Klavans, R.: Classification of
  individual articles from all of science by research level.
\newblock Journal of Informetrics \textbf{8}(1), 1--12 (2014).
\newblock \doi{10.1016/j.joi.2013.10.005}

\bibitem{boyack_improving_2013}
Boyack, K.W., Small, H., Klavans, R.: Improving the accuracy of co-citation
  clustering using full text: {Improving} the {Accuracy} of {Co}-citation
  {Clustering} {Using} {Full} {Text}.
\newblock Journal of the American Society for Information Science and
  Technology \textbf{64}(9), 1759--1767 (2013).
\newblock \doi{10.1002/asi.22896}.
\newblock \urlprefix\url{http://doi.wiley.com/10.1002/asi.22896}

\bibitem{chakraborty_role_2018}
Chakraborty, T.: Role of interdisciplinarity in computer sciences:
  quantification, impact and life trajectory.
\newblock Scientometrics \textbf{114}(3), 1011--1029 (2018).
\newblock \doi{10.1007/s11192-017-2628-z}.
\newblock \urlprefix\url{https://doi.org/10.1007/s11192-017-2628-z}

\bibitem{wos_ref}
{Clarivate Analytics}: {Web of Science} (2019).
\newblock
  \urlprefix\url{https://clarivate.com/webofsciencegroup/solutions/web-of-science/}.
\newblock Accessed Dec 2019

\bibitem{graclus_2007}
Dhillon, I., Guan, Y., Kulis, B.: Weighted graph cuts without eigenvectors: A
  multilevel approach.
\newblock In: {IEEE Transactions on Pattern Analysis and Machine Intelligence
  (PAMI)}, vol. 29:11, pp. 1944--1957. ACM Press (2007)

\bibitem{scopus_ref}
{Elsevier}: {Scopus} (2019).
\newblock \urlprefix\url{https://www.scopus.com/home.uri}.
\newblock Accessed Dec 2019

\bibitem{emmons2016analysis}
Emmons, S., Kobourov, S., Gallant, M., B{\"o}rner, K.: Analysis of network
  clustering algorithms and cluster quality metrics at scale.
\newblock PloS one \textbf{11}(7), e0159161 (2016)

\bibitem{glanzel_using_2017}
Glänzel, W., Thijs, B.: Using hybrid methods and ‘core documents’ for the
  representation of clusters and topics: the astronomy dataset.
\newblock Scientometrics \textbf{111}(2), 1071--1087 (2017).
\newblock \doi{10.1007/s11192-017-2301-6}.
\newblock \urlprefix\url{https://doi.org/10.1007/s11192-017-2301-6}

\bibitem{kessler_comparison_1965}
Kessler, M.M.: Comparison of the results of bibliographic coupling and analytic
  subject indexing.
\newblock American Documentation \textbf{16}(3), 223--233 (1965).
\newblock \doi{10.1002/asi.5090160309}.
\newblock
  \urlprefix\url{https://onlinelibrary.wiley.com/doi/abs/10.1002/asi.5090160309}

\bibitem{klavans_which_2017}
Klavans, R., Boyack, K.W.: Which {Type} of {Citation} {Analysis} {Generates}
  the {Most} {Accurate} {Taxonomy} of {Scientific} and {Technical} {Knowledge}?
\newblock Journal of the Association for Information Science and Technology
  \textbf{68}(4), 984--998 (2017).
\newblock \doi{10.1002/asi.23734}

\bibitem{GithubERNIE2019}
Korobskiy, D., Davey, A., Liu, S., Devarakonda, S., Chacko, G.: {Enhanced
  Research Network Informatics Environment (ERNIE)}.
\newblock Github repository, NET ESolutions Corporation (2019).
\newblock \urlprefix\url{https://github.com/NETESOLUTIONS/ERNIE}

\bibitem{marshakova-shaikevich_co-citation_1973}
Marshakova-Shaikevich, I.: System of document connections based on references.
\newblock Nauchno-Tekhnicheskaya Informatsiya Seriya 2-Informatsionnye
  Protsessy I Sistemy \textbf{6}(4), 3--8 (1973).
\newblock \doi{10.1002/asi.4630240406}

\bibitem{nas_2017}
{National Academies of Sciences, Engineering, and Medicine}, et~al.: Assessing
  and Responding to the Growth of Computer Science Undergraduate Enrollments.
\newblock The National Academies Press, Washington, DC (2018).
\newblock \doi{10.17226/24926}

\bibitem{nsf_classification}
{National Science Foundation}: {Classification of Fields of Study} (2012).
\newblock
  \urlprefix\url{https://www.nsf.gov/statistics/nsf13327/pdf/tabb1.pdf}.
\newblock Accessed June 2019

\bibitem{perianes-rodriguez_comparison_2017}
Perianes-Rodriguez, A., Ruiz-Castillo, J.: A comparison of the {Web} of
  {Science} and publication-level classification systems of science.
\newblock Journal of Informetrics \textbf{11}, 32--45 (2017).
\newblock \doi{10.1016/j.joi.2016.10.007}

\bibitem{salton_citation_1979}
Salton, G., Bergmark, D.: A citation study of computer science literature.
\newblock IEEE Transactions on Professional Communication \textbf{PC-22}(3),
  146--158 (1979).
\newblock \doi{10.1109/TPC.1979.6501740}

\bibitem{shu_comparing_2019}
Shu, F., Julien, C.A., Zhang, L., Qiu, J., Zhang, J., Larivière, V.: Comparing
  journal and paper level classifications of science.
\newblock Journal of Informetrics \textbf{13}(1), 202--225 (2019).
\newblock \doi{10.1016/j.joi.2018.12.005}

\bibitem{shun_parallel_2016}
Shun, J., Roosta-Khorasani, F., Fountoulakis, K., Mahoney, M.W.: Parallel
  {Local} {Graph} {Clustering}.
\newblock Proc. VLDB Endow. \textbf{9}(12), 1041--1052 (2016).
\newblock \doi{10.14778/2994509.2994522}

\bibitem{siebel2019_digital}
Siebel, T.: Digital transformation: survive and thrive in an era of mass
  extinction.
\newblock RosettaBooks (2019)

\bibitem{sjogarde_granularity_2019}
Sjögårde, P., Ahlgren, P.: Granularity of algorithmically constructed
  publication-level classifications of research publications: {Identification}
  of specialties.
\newblock Quantitative Science Studies pp. 1--32 (2019).
\newblock \doi{{10.1162/qss_a_00004}}.
\newblock \urlprefix\url{{https://doi.org/10.1162/qss_a_00004}}

\bibitem{small_co-citation_1973}
Small, H.: Co-citation in the scientific literature: {A} new measure of the
  relationship between two documents.
\newblock Journal of the American Society for Information Science
  \textbf{24}(4), 265--269 (1973).
\newblock \doi{10.1002/asi.4630240406}

\bibitem{small_structure_1974}
Small, H., Griffith, B.C.: The {Structure} of {Scientific} {Literatures} {I}:
  {Identifying} and {Graphing} {Specialties}.
\newblock Science Studies \textbf{4}(1), 17--40 (1974).
\newblock \doi{10.1177/030631277400400102}

\bibitem{small_clustering_1985}
Small, H., Sweeney, E.: Clustering the science citation index using
  co-citations.
\newblock Scientometrics \textbf{7}(3), 391--409 (1985).
\newblock \doi{10.1007/BF02017157}

\bibitem{dblp_ref}
{The dblp Team}: {dblp Computer Science Bibliography} (2018).
\newblock \urlprefix\url{https://dblp.org/xml/release/dblp-2018-08-01.xml.gz}.
\newblock Accessed June 2019

\bibitem{traag_louvain_2019}
Traag, V.A., Waltman, L., {van Eck}, N.J.: From {Louvain} to {Leiden}:
  guaranteeing well-connected communities.
\newblock Scientific Reports \textbf{9}(1), 1--12 (2019).
\newblock \doi{10.1038/s41598-019-41695-z}

\bibitem{subelj_clustering_2016}
\v{S}ubelj, L., {van Eck}, N.J., Waltman, L.: Clustering {Scientific}
  {Publications} {Based} on {Citation} {Relations}: {A} {Systematic}
  {Comparison} of {Different} {Methods}.
\newblock PLOS ONE \textbf{11}(4), e0154404 (2016).
\newblock \doi{10.1371/journal.pone.0154404}

\bibitem{waltman_new_2012}
Waltman, L., {van Eck}, N.J.: A new methodology for constructing a
  publication-level classification system of science.
\newblock Journal of the American Society for Information Science and
  Technology \textbf{63}(12), 2378--2392 (2012).
\newblock \doi{10.1002/asi.22748}

\bibitem{wang_large-scale_2016}
Wang, Q., Waltman, L.: Large-scale analysis of the accuracy of the journal
  classification systems of {Web} of {Science} and {Scopus}.
\newblock Journal of Informetrics \textbf{10}(2), 347--364 (2016).
\newblock \doi{10.1016/j.joi.2016.02.003}

\end{thebibliography}


\end{document}